\documentclass{jpconf}

\usepackage{graphicx}
\usepackage{amsmath,amssymb,bm}
\usepackage{cite}

\begin{document}

\title{Nonlinear Schr\"odinger equations: Symmetries, superposition, and classicality from a Bohmian perspective}

\author{\'Angel S. Sanz}

\address{Department of Optics, Faculty of Physical Sciences, Universidad Complutense de Madrid\\ Pza.\ Ciencias 1, 28040 Madrid, Spain}


\ead{a.s.sanz@fis.ucm.es}


\begin{abstract}
Interference is commonly regarded as the most direct manifestation of the superposition principle. This association is natural for the linear Schr\"odinger equation, where coherent alternatives combine at the level of probability amplitudes. However, the situation becomes less transparent when nonlinear couplings are present, or when the field is only partially coherent. In this work, we argue that a more robust organizing principle is provided by the local flow generated by phase variations. In this sense, phase-induced flow acts as a unifying mechanism for interference-like dynamics in nonlinear and partially coherent Schr\"odinger systems. The discussion is developed from a hydrodynamic, or Bohmian, perspective, understood here as a practical probing tool rather than as an additional ontology. Three representative situations are considered: interfering Bose--Einstein condensates described by the Gross--Pitaevskii equation, nonlinear Schr\"odinger dynamics obtained by modifying the quantum-potential contribution, and partially coherent Airy beams described through their cross-spectral density. Although these systems differ in physical origin and mathematical implementation, they share a common dynamical structure: density-related observables are shaped by velocity fields determined by phase, or ensemble-phase, information. From this viewpoint, interference-like traits, localization, self-acceleration and coherence loss can be interpreted in terms of the preservation, deformation or breaking of the symmetries displayed by the underlying flow. This provides a compact way of connecting interference, nonlinear dynamics, classicality, coherence loss, and structured-light propagation within a single trajectory-based framework.
\end{abstract}


\section{Introduction}

Interference is often presented as the most characteristic signature of the superposition principle. In the linear Schr\"odinger equation, this connection is direct: two coherent alternatives, represented by $\psi_A$ and $\psi_B$, are added at the amplitude level, and the corresponding density contains the well-known cross terms.
This simple algebraic fact underlies some paradigmatic manifestations of quantum coherence, from the buildup of single-particle interference patterns \cite{Tonomura1989,arndt:Nature:1999,arndt:NaturePhys:2014} to the observation of interference between independently prepared Bose--Einstein condensates \cite{ketterle:Science:1997,schumm:NaturePhys:2005}. It also lies behind the operational reconstruction of average photon trajectories in two-slit interferometry by means of weak measurements \cite{Kocsis2011}, and, more broadly, the use of coherence as a resource in coherent control \cite{brumer-bk:2003} and quantum-information-based science and technologies \cite{nielsen-chuang-bk}.

However, interference is not exhausted by the algebraic statement that amplitudes can be added. Once the wave function is written in polar form,
\begin{equation}
 \psi(\bm r,t)=A(\bm r,t)\exp\left[\frac{i S(\bm r,t)}{\hbar}\right],
 \label{eq:polar_intro}
\end{equation}
one can also define the local velocity field
\begin{equation}
 \bm v(\bm r,t)=\frac{\bm j(\bm r,t)}{\rho(\bm r,t)}
 =\frac{\hbar}{m}\operatorname{Im}\left[\frac{\nabla\psi(\bm r,t)}{\psi(\bm r,t)}\right]
 =\frac{\nabla S(\bm r,t)}{m},
 \label{eq:velocity_intro}
\end{equation}
where $\rho=|\psi|^2$ and $\bm j=\rho\bm v$. The density satisfies the continuity equation
\begin{equation}
 \frac{\partial\rho}{\partial t}+\nabla\cdot\bm j=0,
 \label{eq:continuity_intro}
\end{equation}
which shows that the observed distribution is the result of an underlying transport process. From this viewpoint, interference is not only a pattern in the density, but also a phase-induced redistribution of probability flux.
The symmetries displayed by this flux, or their breaking, are then directly reflected in the observable pattern: preserving a phase-flow symmetry sustains robust structures, whereas breaking it leads to deformation, localization, or loss of contrast.

This is precisely the aspect emphasized by the hydrodynamic formulation of quantum mechanics. Since its formulation by Bohm in 1952 \cite{Bohm1952a,Bohm1952b}, Bohmian mechanics has often been regarded as a causal interpretation of quantum mechanics \cite{Holland1993}, or as a theory with its own ontology.
Yet, from a strictly formal viewpoint, the same equations also arise as a direct consequence of rewriting the Schr\"odinger equation in hydrodynamic terms, as Madelung had already shown in 1926 \cite{Madelung1926}. In this latter sense, which is the one adopted here, Bohmian mechanics does not need to be invoked as a hidden-variable theory. It can simply be used as a convenient language to display, in local terms, the information already encoded in the wave function, particularly the information associated with the phase \cite{SanzFront2019,SanzJPCS2024}.

This distinction between global and local information is important.
The wave function, or its associated density, provides the usual global description of the physical state. However, the way in which such a density redistributes in configuration space is governed by a current, or flux, and hence by the associated velocity field \cite{SanzFront2019}.
This velocity field is directly related to the local spatial variations undergone by the phase. Accordingly, the so-called Bohmian trajectories may be understood, leaving aside ontological connotations, as streamlines of a well-defined flux. In the pragmatic use considered here, they are not real paths followed by a quantum particle, but a local probing tool that allows us to visualize how probability or optical intensity is transported \cite{SanzJPCS2024}.

The situation becomes subtler when the superposition principle is not available in its standard form. Nonlinear Schr\"odinger equations (NSEs) do not allow arbitrary linear combinations of solutions. Nevertheless, they may display interference-like traits, phase-induced localization, focusing, or the persistence of structured propagation. Something analogous happens in partially coherent optics \cite{mandelwolf-bk}. There, the field is no longer described by a single pure amplitude, but by an ensemble quantity such as the cross-spectral density. Even so, as shown in \cite{MartinezSanz2025}, a current and an associated flux velocity can still be defined. Thus, the relevant question is not only whether amplitudes can be superposed, but whether phase information, or its ensemble counterpart, continues to organize the local dynamics.

The purpose here is to discuss this idea in a compact and unified manner. To this end, three representative examples, drawn from previous analyses and associated with different physical scenarios, are considered. First, we analyze interfering Bose--Einstein condensates governed by the Gross--Pitaevskii equation, where the nonlinearity enters as a mean-field contribution proportional to the density \cite{Gross1961,Pitaevskii1961,PethickSmith2002,TounliSanz2024}. Second, we discuss a nonlinear ``classical'' Schr\"odinger equation in which the contribution of Bohm's quantum potential is gradually suppressed \cite{Schiller1962a,Schiller1962b,Rosen1964,Rosen1986,NaviaSanz2024}. Third, we move beyond pure states and consider partially coherent Airy beams, whose propagation can be described in terms of generalized flux trajectories obtained from the cross-spectral density \cite{BerryBalazs1979,Siviloglou2007,MartinezSanz2022,SanzMartinez2024,MartinezSanz2025}.

The common message is that phase-induced flow provides a useful unifying mechanism for understanding interference-like dynamics in nonlinear and partially coherent Schr\"odinger systems. In this framework, the relevant symmetries are those displayed by the phase and velocity fields, since they determine how probability or optical intensity is locally transported. The systems considered below are not physically equivalent; rather, they can be compared through the same dynamical object: a velocity field that locally accounts for the transport of probability or optical intensity.

The work is organized as follows. In Sec.~\ref{sec:phaseflow}, the basic hydrodynamic formulation and its paraxial-optical counterpart are summarized. Section~\ref{sec:interference} discusses ordinary coherent superpositions as phase-driven flux redistributions. Section~\ref{sec:gpe} considers the Gross--Pitaevskii equation. Section~\ref{sec:classical_sch} analyzes the ``classical'' Schr\"odinger equation and the relation between quantum-potential cancellation and classicality. Section~\ref{sec:partial_coherence} extends the discussion to partially coherent structured beams. Some concluding remarks are given in Sec.~\ref{sec:conclusions}.


\section{Phase-induced flow in Schr\"odinger-type systems}
\label{sec:phaseflow}

Let us consider, for simplicity, a spinless particle with nonzero mass, described by the time-dependent Schr\"odinger equation
\begin{equation}
 i\hbar\frac{\partial\psi(\bm r,t)}{\partial t}
 = -\frac{\hbar^2}{2m}\nabla^2\psi(\bm r,t) + V(\bm r,t)\psi(\bm r,t).
 \label{eq:linear_sch}
\end{equation}
Multiplying Eq.~(\ref{eq:linear_sch}) by $\psi^*$, subtracting the complex conjugate equation, and rearranging the result, one obtains the continuity equation introduced above, Eq.~(\ref{eq:continuity_intro}), with the probability current
\begin{equation}
 \bm j(\bm r,t)=\frac{\hbar}{2mi}
 \left[ \psi^*(\bm r,t)\nabla\psi(\bm r,t) - \psi(\bm r,t)\nabla\psi^*(\bm r,t) \right] ,
 \label{eq:current}
\end{equation}
where $V(\bm r,t)$ may depend on time, as in typical light--matter interaction problems.
Since this is a transport equation, the current can be written as $\bm j=\rho\bm v$, which leads to the velocity field in Eq.~(\ref{eq:velocity_intro}). Equivalently, the associated local momentum field is
\begin{equation}
 {\bm p}({\bm r},t) = m{\bm v}({\bm r},t) = 
 {\rm Re}\left[ \frac{\hat{\bm p}\psi({\bm r},t)}{\psi({\bm r},t)} \right],
 \qquad \hat{\bm p}\equiv-i\hbar\nabla .
 \label{eq:local_momentum}
\end{equation}
Thus, the quantity that generates the local flow is not external to standard quantum mechanics. It is already encoded in the wave function through the local action of the momentum operator.

A complementary and more geometrical view is obtained by substituting the polar form (\ref{eq:polar_intro}) into Eq.~(\ref{eq:linear_sch}) and separating real and imaginary parts \cite{Bohm1952a,Holland1993}. The imaginary part gives
\begin{equation}
 \frac{\partial\rho({\bm r},t)}{\partial t}
 + \nabla\cdot\left[\rho({\bm r},t)\frac{\nabla S({\bm r},t)}{m}\right]=0,
 \label{eq:continuity_phase}
\end{equation}
whereas the real part yields the quantum Hamilton--Jacobi equation
\begin{equation}
 \frac{\partial S({\bm r},t)}{\partial t} +\frac{\left[\nabla S({\bm r},t)\right]^2}{2m}
 +V({\bm r},t) + Q({\bm r},t)=0,
 \label{eq:qhje}
\end{equation}
with
\begin{equation}
 Q(\bm r,t) = -\frac{\hbar^2}{2m} \frac{\nabla^2 A(\bm r,t)}{A(\bm r,t)}
  = \frac{\hbar^2}{8m} \left\{ \left[\frac{\nabla \rho(\bm r,t)}{\rho(\bm r,t)}\right]^2 - 2 \frac{\nabla^2 \rho(\bm r,t)}{\rho(\bm r,t)} \right\}
 \label{eq:quantum_potential}
\end{equation}
being Bohm's quantum potential. This term should not be understood as an external potential energy, but as the amplitude-curvature contribution arising from the kinetic operator.
In the hydrodynamic picture, the evolution is therefore described in
terms of two intertwined pieces of information: the phase, through
$\nabla S(\bm r,t)$, and the amplitude, through $Q(\bm r,t)$.

The integral curves of the velocity field,
\begin{equation}
 \frac{d{\bm r}}{dt} = {\bm v}({\bm r},t) = \frac{\nabla S(\bm r,t)}{m},
 \label{eq:guidance}
\end{equation}
provide the hydrodynamic streamlines of the probability flux. This is the sense in which the term Bohmian trajectories will be used below. They offer a local representation of the flow associated with the wave field, without requiring any additional ontological commitment.
In this representation, symmetries of the phase field are inherited by the velocity field and, consequently, by the corresponding family of streamlines.

It is also worth noting that quantities closely related to the local momentum field can be accessed operationally through weak-measurement schemes. This is why average trajectories reconstructed in interferometric experiments can be meaningfully compared with the streamlines obtained from Eq.~(\ref{eq:guidance}) \cite{Kocsis2011}.
In this respect, the hydrodynamic formulation is not only a formal rewriting, but also a useful language to connect density-related information with phase-related information.

The same formal structure appears in scalar paraxial optics \cite{sanz-bk}. If a monochromatic field mainly propagates along the longitudinal coordinate $z$, its slowly varying transverse amplitude satisfies
\begin{equation}
 i\frac{\partial\psi(\bm r_\perp,z)}{\partial z}
 =
 -\frac{1}{2k}\nabla_\perp^2\psi(\bm r_\perp,z),
 \label{eq:paraxial_basic}
\end{equation}
which is formally equivalent to a free Schr\"odinger equation. The propagation coordinate $z$ plays the role of the evolution variable, whereas the relevant dynamics takes place in the transverse plane. The corresponding effective transverse velocity field is
\begin{equation}
 \frac{d\bm r_\perp}{dz}
 =
 \bm v(\bm r_\perp,z)
 =
 \frac{1}{k}
 {\rm Re}\left[
 \frac{-i\nabla_\perp\psi(\bm r_\perp,z)}
 {\psi(\bm r_\perp,z)}
 \right].
 \label{eq:paraxial_velocity_basic}
\end{equation}
Although this is not a mechanical velocity, it describes how the transverse distribution of optical intensity is locally redistributed during propagation. This analogy allows us to discuss matter waves and structured light within the same phase-flow language.
The next step is to see how this local flow behaves when the wave field is built from coherent alternatives.


\section{Interference as phase-driven flux redistribution}
\label{sec:interference}

Before moving to explicitly nonlinear equations, it is useful to recall that
ordinary coherent superpositions already contain a nontrivial hydrodynamic
structure. Let us consider a two-component wave field,
\begin{equation}
 \psi (\bm r,t)=\psi_1(\bm r,t) + \psi_2(\bm r,t) ,
 \qquad
 \psi_\beta(\bm r,t) = A_\beta(\bm r,t) \exp\left[\frac{i S_\beta(\bm r,t)}{\hbar}\right] ,
 \label{eq:twocomp}
\end{equation}
with $\beta=1,2$.
The corresponding density is
\begin{equation}
 \rho(\bm r,t)=\rho_1(\bm r,t)+\rho_2(\bm r,t)+2\sqrt{\rho_1(\bm r,t)\rho_2(\bm r,t)}\cos\varphi(\bm r,t) ,
 \label{eq:twocomp_density}
\end{equation}
with $\rho_\beta(\bm r,t)=A_\beta^2(\bm r,t)$ and relative phase
\begin{equation}
\varphi(\bm r,t)= [S_1(\bm r,t)-S_2(\bm r,t)]/\hbar .
\end{equation}
The third contribution on the right-hand side of Eq.~\eqref{eq:twocomp_density} is the usual interference term. However, the hydrodynamic content of
the superposition is not limited to this density modulation. The associated
current reads
\begin{align}
 \bm j(\bm r,t)
 =&\frac{1}{m}
 \left[
 \rho_1(\bm r,t)\nabla S_1(\bm r,t)
 +
 \rho_2(\bm r,t)\nabla S_2(\bm r,t)
 \right]
 \nonumber\\
 &+
 \frac{\sqrt{\rho_1(\bm r,t)\rho_2(\bm r,t)}}{m}
 \left[
 \nabla S_1(\bm r,t)
 +
 \nabla S_2(\bm r,t)
 \right]
 \cos\varphi(\bm r,t)
 \nonumber\\
 &+
 \frac{\hbar}{2m}
 \sqrt{\rho_1(\bm r,t)\rho_2(\bm r,t)}
 \left[
 \frac{\nabla\rho_1(\bm r,t)}{\rho_1(\bm r,t)}
 -
 \frac{\nabla\rho_2(\bm r,t)}{\rho_2(\bm r,t)}
 \right]
 \sin\varphi(\bm r,t).
 \label{eq:twocomp_current}
\end{align}
Thus, although Eq.~(\ref{eq:twocomp}) is a linear superposition at the level
of amplitudes, the velocity field $\bm v=\bm j/\rho$ is a nonlinear
functional of the component amplitudes and phases.

This observation is central for the present discussion. The superposition
principle explains why the amplitude can be written as a sum, but the local
dynamics is governed by the velocity field generated by that sum. In other
words, interference is not only a matter of adding amplitudes; it also
involves the organization of a global flux field. This field determines how
probability or intensity is redistributed in configuration space, giving
rise to the non-crossing of streamlines and to the channel-like structures
that characterize interference dynamics \cite{SanzMiret2008}.
These channel-like structures can be understood as manifestations of a dynamical symmetry imposed by the global phase field: trajectories launched in different regions remain organized by the same velocity field, even though the final density only displays maxima and minima.

\begin{figure}[t]
\centering
\includegraphics[width=\textwidth]{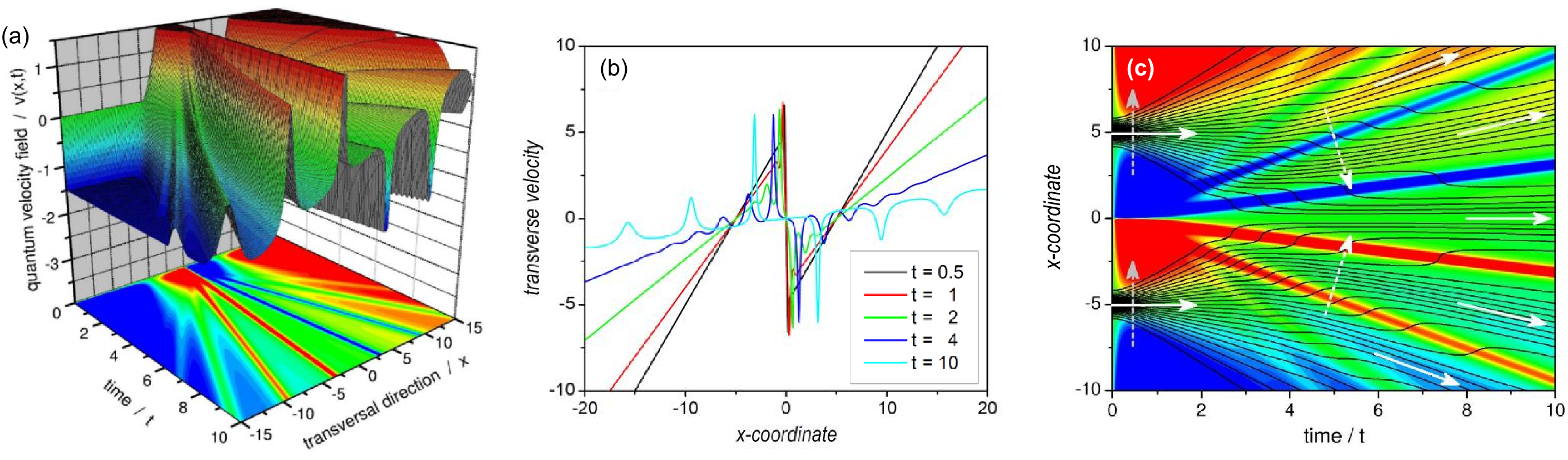}
\caption{Phase-induced flow information in a Young-type interference scenario.
The panels illustrate complementary aspects of the same wave dynamics:
(a) evolution in time of the quantum velocity field $v(x,t)$, (b) variation of the velocity field along the $x$-coordinate at different specific times, and (c) Bohmian trajectories superimposed on a density plot of the velocity field.
In panel (c), gray arrows indicate increasing values of the velocity field, while white arrows denote the average local motion exhibited by the trajectories.
Note that, while a typical density distribution provides the observable interference structure, the associated velocity field or flux trajectories reveal the local redistribution of probability induced by phase variations. This distinction is essential for the discussion below, where analogous phase-induced flows persist even when the standard superposition principle is no longer directly applicable.}
\label{fig1}
\end{figure}

A convenient way of visualizing this distinction is shown in
Fig.~\ref{fig1}.
A typical plot of the density distribution provides a visual representation of how different spatial regions get populated over time, directly related to the experimentally observable interference pattern; the corresponding velocity field or flux trajectories, on the other hand, reveal how that pattern is dynamically built from the local phase structure: interference maxima are associated with plateau-like regions in the velocity field, with relatively high accumulations of trajectories.
This density/phase-flow complementarity is the key point to retain before entering the nonlinear cases.

This is the point that will be exploited below. Once nonlinear couplings are present, the superposition principle cannot be appealed to in the usual way. However, the phase still generates a velocity field and this field still determines how the density is locally redistributed.
Therefore, rather than asking whether an interference pattern is compatible with nonlinearity, it is more convenient to ask how the nonlinear evolution modifies the phase-induced flow that gives rise to interference-like structures.


\section{Interference-like dynamics in the Gross--Pitaevskii equation}
\label{sec:gpe}

Matter-wave interferometry with Bose--Einstein condensates provides a natural scenario for the present discussion, because phase coherence becomes visible at a macroscopic scale \cite{ketterle:Science:1997,schumm:NaturePhys:2005}.
At the same time, after release, the corresponding mean-field dynamics is no longer governed by a linear Schr\"odinger equation, but by the Gross--Pitaevskii equation,
\begin{equation}
 i\hbar \frac{\partial \psi(z,t)}{\partial t}
 =
 -\frac{\hbar^2}{2m}\frac{\partial^2\psi(z,t)}{\partial z^2}
 +g|\psi(z,t)|^2\psi(z,t),
 \label{eq:gpe}
\end{equation}
where the nonlinear contribution can be understood as the density-dependent
mean-field potential
\begin{equation}
 V_{\rm mf}(z,t)=g|\psi(z,t)|^2.
 \label{eq:vmf}
\end{equation}
For the purpose of the present discussion, we focus on the free expansion and mutual overlap of two initially separated condensate fragments. Once released, their dynamics is affected only by the nonlinear mean-field term.
A simple way of representing the initial condition is
\begin{equation}
 \psi(z,0)\sim
 \exp\left[-\frac{(z-z_+)^2}{4\sigma_0^2}\right]
 +e^{i\phi}
 \exp\left[-\frac{(z-z_-)^2}{4\sigma_0^2}\right].
 \label{eq:gpe_initial_round4}
\end{equation}
The relevant parameters are then the initial size of the two components,
their mutual overlap, and the imprinted phase difference $\phi$; further
details on the specific Gross--Pitaevskii simulations can be found in
Ref.~\cite{TounliSanz2024}.

From the density viewpoint, the merging of the two condensate fragments
gives rise to dark structures whose number, contrast and subsequent
motion depend on those ingredients. From the phase-flow viewpoint, the
same process is more naturally understood as a redistribution of flux
generated by the local phase profile and modified by the nonlinear
mean-field term. The nonlinearity changes the phase evolution, and the
phase determines the current.
Accordingly, any symmetry initially present in the relative phase profile, or its breaking through an imprinted phase difference, becomes dynamically visible in the subsequent flux pattern.
Hence, the dark structures are not merely
static minima associated with destructive interference, but dynamical
features embedded in a nonlinear flow.

\begin{figure}[t]
\centering
\includegraphics[width=\textwidth]{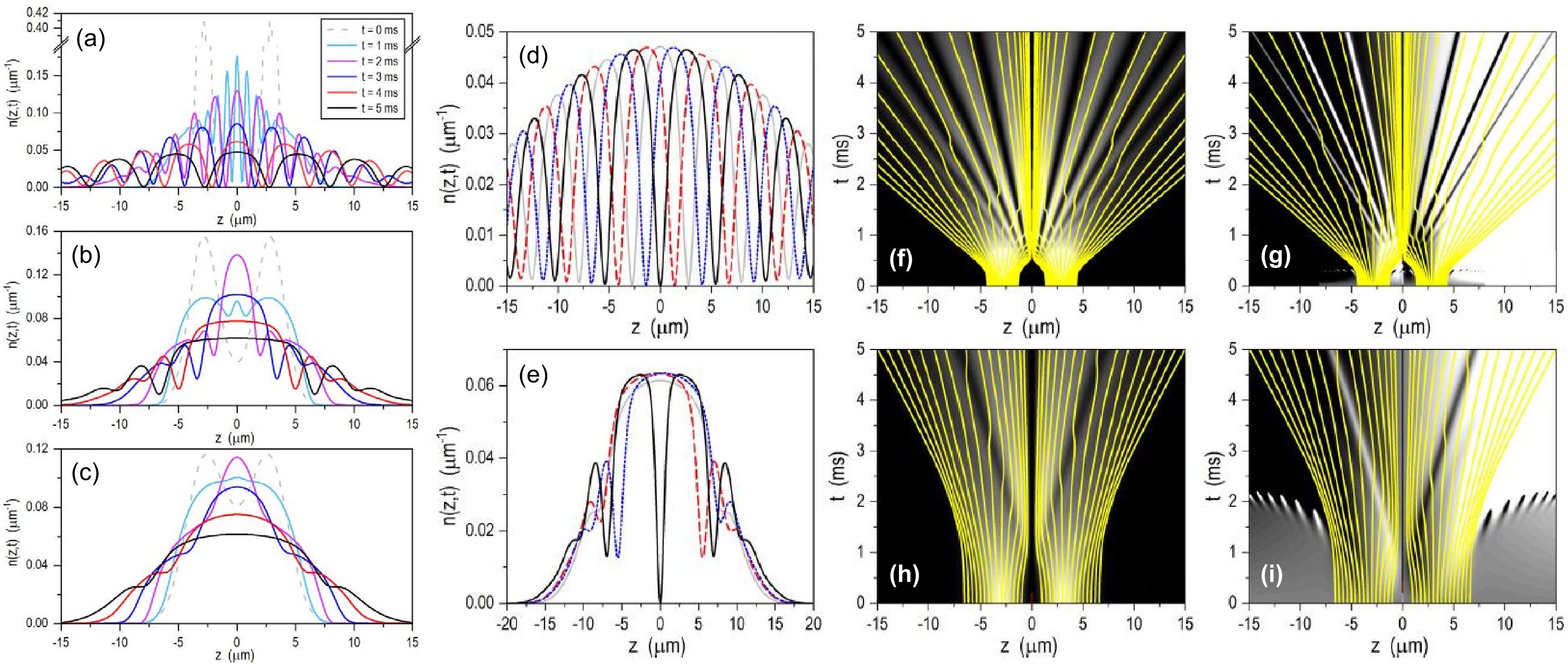}
\caption{Gross--Pitaevskii dynamics for two initially separated condensate fragments.
Density evolution for different values of the overlapping parameter $S \equiv |\psi(0,0)|^2/|\psi(z_\pm,0)|^2$: (a) $S \approx 0$, (b) $S \approx 0.26$, and (c) $S \approx 0.7$.
Density structure displayed at $t = 5$~ms by the wave packets with $S \approx 0$ (d) and $S \approx 0.7$ (e) depending on the relative phase $\phi$ imprinted between both components at $t=0$: 0 (gray solid line), $\pi/2$ (red dashed line), $\pi$ (black solid line), and $-\pi/2$ (blue dotted line).
The flux trajectories for $S\simeq 0$ with $\phi=0$ and $\phi=\pi$ are represented, respectively, in panels (f,g) and (h,i).
For each phase value, the shaded density plot represents the density distribution in panels (f,h), and the velocity field in panels (g,i).
Note from the hydrodynamic viewpoint that the emerging dark structures are not merely interference minima, but dynamical features sustained by the velocity field generated by the local phase profile and modified by the nonlinear mean-field coupling.}
\label{fig2}
\end{figure}

This behavior is illustrated in Fig.~\ref{fig2}. The comparison between
different initial conditions shows that the formation of dark structures
is controlled not only by the overlap between the two wave packets, but
also by the relative phase imprinted between them. In particular, changing
$\phi$ modifies the initial phase landscape without necessarily producing
a large change in the initial density. The subsequent dynamics can then
be substantially different because the local velocity field has changed.

In this sense, the Gross--Pitaevskii scenario shows that nonlinearity does
not erase the phase-flow picture.
It rather makes this picture even more necessary: the density alone tells us where the dark structures appear, but the velocity field tells us how they are generated and transported.


\section{Classicality beyond quantum-potential cancellation}
\label{sec:classical_sch}

The Gross--Pitaevskii equation provides an example in which the nonlinearity has a clear mean-field origin.
Let us now consider a different situation in which the nonlinear term is introduced with a more conceptual purpose: to modify the role played by Bohm's quantum potential in the Hamilton--Jacobi equation.
\begin{equation}
 i\hbar\frac{\partial\psi(\bm r,t)}{\partial t}
 =
 -\frac{\hbar^2}{2m}\nabla^2\psi(\bm r,t)
 +V(\bm r)\psi(\bm r,t)
 +\gamma\frac{\hbar^2}{2m}
 \frac{\nabla^2A(\bm r,t)}{A(\bm r,t)}
 \psi(\bm r,t).
 \label{eq:classical_sch_round4}
\end{equation}
The parameter $\gamma$ controls the strength of this nonlinear
correction. For $\gamma=0$, Eq.~(\ref{eq:classical_sch_round4}) reduces
to the usual Schr\"odinger equation.
For $\gamma=1$, the contribution associated with the quantum potential is canceled in the Hamilton--Jacobi equation, so that the phase obeys a classical-like Hamilton--Jacobi dynamics \cite{Schiller1962a,Schiller1962b,Rosen1964,Rosen1986,NaviaSanz2024}.

At first sight, this construction seems to provide a direct route from quantum to classical dynamics. If the quantum potential is the term that distinguishes Eq.~(\ref{eq:qhje}) from the classical Hamilton--Jacobi equation, then suppressing it could be expected to suppress the quantum behavior. This expectation, however, overlooks the role played by coherence. The nonlinear term changes the equation for the phase, but it does not automatically erase the phase relations already encoded in the wave field.
Thus, flow symmetries inherited from the original coherent phase field may survive even when the explicit quantum-potential contribution has been canceled.
As a consequence, the resulting evolution may still display features that are not those of a classical statistical ensemble.

This point is important for the present discussion. Classicality is not
only a matter of canceling a formal term from the equation of motion. It
also involves the disappearance, or at least the dynamical irrelevance, of
the organized phase relations that sustain wave-like flow. Therefore,
localization, reduced spreading, or focusing-like behavior should not be
identified by themselves with a full quantum-to-classical transition. The
velocity field remains the diagnostic object: if it preserves
coherence-induced structure, the dynamics still carries nonclassical
information.

\begin{figure}[t]
\centering
\includegraphics[width=\textwidth]{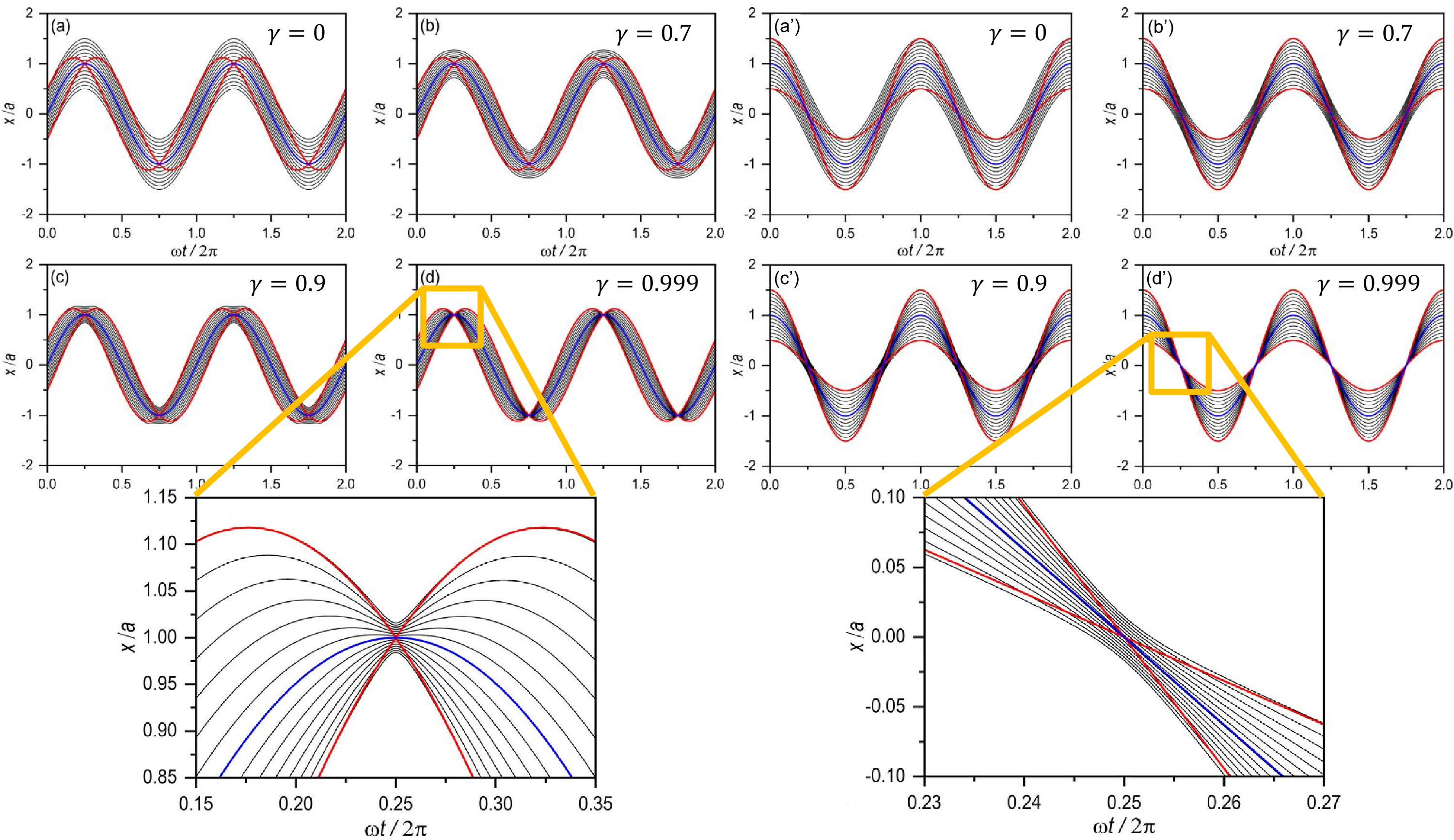}
\caption{Dynamics generated by the nonlinear ``classical'' Schr\"odinger
equation for different values of the coupling parameter $\gamma$ in a coherent wave packet oscillating inside a harmonic potential.
Black solid lines indicate flux trajectories with different initial conditions inside the well; the two red solid lines represent classical trajectories starting with the same initial positions as the marginal flux trajectories; and the blue solid line denotes the classical trajectory starting at the center of the wave packet, which also describes the behavior of the position expectation value $\langle\hat{x}\rangle(t)$.
Panels (a)--(d) represent a situation where motion starts at the center of the wave packet, while panels (a')--(d') refer to motion started from one of the turning points.
Note that, as $\gamma$ approaches the value for which the quantum-potential contribution is canceled in the Hamilton--Jacobi equation, the evolution displays enhanced localization or focusing-like features in both cases.
Yet the non-crossing structure of the trajectories reflects the fact that the associated velocity field still carries the coherent phase information of the wave field, as shown in the two enlarged panels.
This figure thus illustrates that suppressing the quantum potential is not, by itself, equivalent to reaching a fully classical regime.}
\label{fig3}
\end{figure}

This distinction is illustrated in Fig.~\ref{fig3}. Increasing $\gamma$
may reduce the spreading of the wave packet, or even produce a
focusing-like dynamics, but the underlying flow still contains the
imprint of the phase field. In this sense, the ``classical''
Schr\"odinger equation provides a useful warning: classicality is not
obtained by canceling $Q(\bm r,t)$ alone. It requires that the phase-induced flow
ceases to behave as a coherent wave flow.


\section{From pure states to partially coherent structured beams}
\label{sec:partial_coherence}

The two previous examples still refer to fields described by a single
complex amplitude. Let us now move one step further and consider a
situation where such a pure-state description is no longer appropriate.
This is the case of partially coherent optical fields, for which the
relevant object is not a wave function, but the cross-spectral density
\cite{mandelwolf-bk},
\begin{equation}
 \mathcal{W}(\bm r_\perp',\bm r_\perp,z)
 =
 \sum_n p_n\phi_n^*(\bm r_\perp',z)\phi_n(\bm r_\perp,z),
 \label{eq:csd_round4}
\end{equation}
where the coherent modes $\phi_n$ obey the paraxial equation. The
intensity is obtained from the diagonal elements,
\begin{equation}
 I(\bm r_\perp,z)
 =
 \mathcal{W}(\bm r_\perp,\bm r_\perp,z),
 \label{eq:intensity_round4}
\end{equation}
while the corresponding current can be written as
\begin{equation}
 \bm j(\bm r_\perp,z)
 =
 \frac{1}{2ik}
 \left[
 \nabla_\perp\mathcal{W}(\bm r_\perp',\bm r_\perp,z)
 -
 \nabla_\perp'\mathcal{W}(\bm r_\perp',\bm r_\perp,z)
 \right]_{\bm r_\perp'=\bm r_\perp}.
 \label{eq:generalized_current_round4}
\end{equation}
Accordingly, one can define the generalized flux velocity
\begin{equation}
 \frac{d\bm r_\perp}{dz} =
 \bm v(\bm r_\perp,z) =
 \frac{\bm j(\bm r_\perp,z)}{I(\bm r_\perp,z)}.
 \label{eq:generalized_velocity_round4}
\end{equation}
Equation~(\ref{eq:generalized_velocity_round4}) is the ensemble analogue
of Eq.~(\ref{eq:paraxial_velocity_basic}). The relevant point is that the
local description is not lost when the field ceases to be fully coherent.
What changes is the object from which the flow is computed: the phase of
a single amplitude is replaced by the phase correlations encoded in the
cross-spectral density.

Airy beams provide a particularly useful test case for this idea.
In the fully coherent limit, the ideal Airy solution displays shape invariance and self-acceleration during propagation \cite{BerryBalazs1979}.
In dimensionless variables, the corresponding solution can be written as
\begin{equation}
 \psi(x,z) =  \exp\left[ \frac{iz}{2} \left( x - \frac{z^2}{6} \right) \right]
 {\rm Ai}\left(x-\frac{z^2}{4}\right),
 \label{eq:airy_round4}
\end{equation}
which gives
\begin{equation}
 \frac{dx}{dz}
 =
 \frac{z}{2},
 \qquad
 x(z)=x_0+\frac{z^2}{4}.
 \label{eq:airy_traj_round4}
\end{equation}
Thus, all parts of the beam are displaced according to the same parabolic law. The phase-flow picture provides a simple way of understanding why shape invariance and transverse self-acceleration appear together: both properties arise from the symmetry displayed by the underlying velocity field.
In practice, however, ideal Airy beams cannot be realized with infinite energy content. Their spatial extension has to be effectively truncated \cite{Siviloglou2007}, and this truncation modifies the flow symmetry responsible for the ideal behavior.

For the class of partially coherent Airy beams considered in
Refs.~\cite{MartinezSanz2022,MartinezSanz2025}, this truncation and the
degree of coherence can be incorporated through a suitable cross-spectral
density built from statistically weighted Airy components. Two parameters
then control the relevant physical effects: one fixes the effective
spatial extension of the Airy structure, while the other determines the
degree of coherence. Their combined action decides how much of the ideal
Airy flow survives at the ensemble level.
In this sense, truncation and partial coherence act as mechanisms that deform, or eventually break, the flow symmetry responsible for the ideal Airy behavior.

\begin{figure}[t]
\centering
\includegraphics[width=0.85\textwidth]{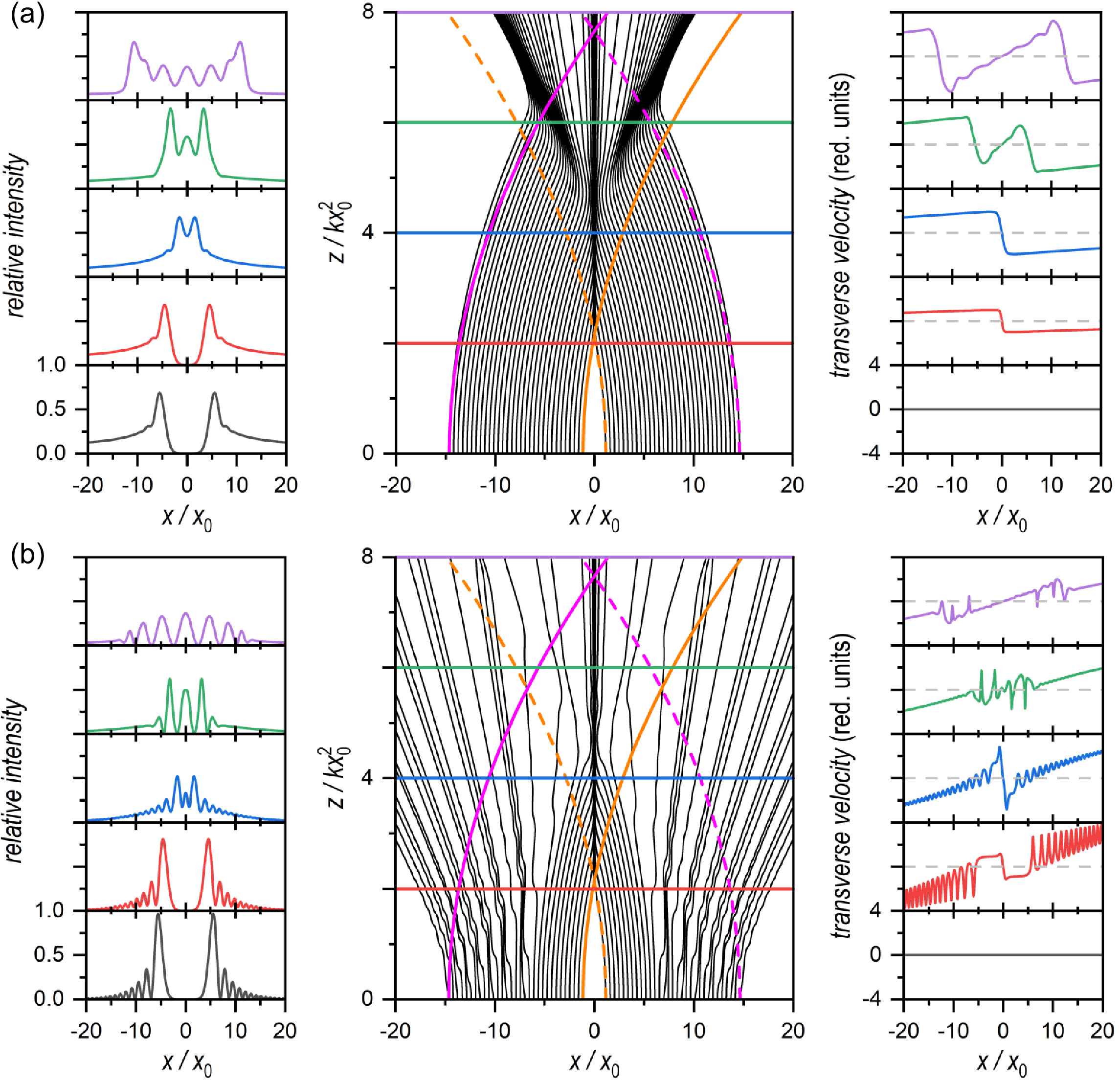}
\caption{Generalized flux trajectories for partially coherent Airy beams.
The panels compare different combinations of spatial extension and degree
of coherence, controlled by the parameters $\sigma$ and $\mu$ in the
cross-spectral density. (a) When the effective Airy tail remains sufficiently
extended ($\sigma = 1,\ \mu = 0.125\sqrt{2}$), the ensemble flow still supports an Airy-like self-accelerating
behavior. (b) As the spatial extension or coherence is reduced ($\sigma = 0.5,\ \mu = 2\sqrt{2}$), the generalized trajectories reveal the gradual smearing of the intensity distribution and the loss of the ideal phase-flow structure.
In both examples, the left and right columns denote the density distribution and the velocity field at specific values of the longitudinal coordinate $z$.
These values are indicated with color horizontal lines in the central panels, where two swarms of flux trajectories are shown, each departing from the region originally occupied by the corresponding beam.
The magenta and orange solid/dashed lines denote the reference ideal-Airy trajectories for the left/right beam.}
\label{fig4}
\end{figure}

As illustrated in Fig.~\ref{fig4}, the generalized trajectories obtained
from Eq.~(\ref{eq:generalized_velocity_round4}) reveal that partial
coherence does not simply blur the intensity distribution. It modifies
the ensemble flow that supports the Airy-like propagation. Thus, the
relevant object is no longer the phase of a single realization, but the
phase correlations encoded in $\mathcal{W}$. This is the natural
extension of the phase-flow picture: from Bohmian streamlines associated
with a pure wave function to generalized flux trajectories associated
with a partially coherent ensemble.

\section{Concluding remarks}
\label{sec:conclusions}

The three examples considered above were not chosen because of their
formal similarity alone, but because they stress different ways in which
the usual superposition-based intuition becomes insufficient. In the
Gross--Pitaevskii equation, the nonlinearity arises from a mean-field
coupling proportional to the density. In the ``classical''
Schr\"odinger equation, the nonlinear term is introduced to modify the
contribution of Bohm's quantum potential and to produce a classical-like
Hamilton--Jacobi equation. In partially coherent optics, the description
is no longer based on a single complex amplitude, but on an ensemble
object, the cross-spectral density. Nevertheless, the three cases point
to the same lesson: the density or intensity distribution is only part of
the story. The local flow that redistributes it carries essential
dynamical information, including the symmetries that organize the
propagation and the ways in which such symmetries are preserved,
deformed, or broken.

This is the reason why the hydrodynamic, or Bohmian, representation is
useful in the present context. It does not add a new ingredient to
Schr\"odinger-type dynamics. Rather, it makes explicit information that
is already encoded in the phase of the wave function, or, in the
partially coherent case, in the correlations contained in the
cross-spectral density. The corresponding velocity field and its
streamlines provide a local description of how probability or optical
intensity is transported. In this sense, the trajectory representation
should be understood here as a probing tool, not as an ontological
requirement.

The analysis also suggests that interference, coherence and classicality
should not be characterized only through density patterns or through the
formal structure of the evolution equation. In linear quantum mechanics,
a coherent superposition produces interference fringes, but also a
nonlinear-looking velocity field. In the Gross--Pitaevskii case,
explicit nonlinear coupling modifies this field without erasing its
phase-driven character. In the nonlinear ``classical'' Schr\"odinger
equation, canceling the quantum-potential contribution is not sufficient
to produce a fully classical behavior if the underlying phase-induced
flow still behaves as a coherent wave flow. Finally, in partially
coherent structured light, the same logic survives at the ensemble level
through the generalized current associated with the cross-spectral
density.

Therefore, the relevant question is not simply whether the superposition
principle holds, whether a nonlinear term is present, or whether visible
fringes persist. The more informative question is what happens to the
phase-induced flow. If this flow remains structured, coherence is still
dynamically active, even if its manifestations are deformed, weakened, or
encoded at the ensemble level. Conversely, any genuine loss of
wave-like behavior should eventually be reflected in the degradation of
that flow structure.

This is the main point that we wished to stress. Phase is not a passive
label attached to a wave function. Through its local variations, it
generates a flow, and this flow determines how probability or intensity
is redistributed. The symmetries of this flow, and the way in which they
are preserved, deformed or broken, provide information that is not
apparent from the density alone. This is why phase-induced flow can be
regarded as a unifying mechanism for interference-like dynamics in
nonlinear and partially coherent Schr\"odinger systems.

\section*{Acknowledgments}

This contribution is based on the talk delivered at Symmetries in Science XX, Bregenz, August 3--8, 2025. The author acknowledges support from project PID2021-127781NB-I00.

\section*{References}


\end{document}